\documentclass[10pt,a4paper,twocolumn,english,pra,aps,showpacs,floatfix,groupedaddress,superscriptaddress]{revtex4}
\usepackage[T1]{fontenc}
\usepackage[latin9]{inputenc}
\usepackage{verbatim}
\usepackage{amsmath}
\usepackage{amssymb}
\usepackage{graphicx}
\usepackage{esint}

\makeatletter

\pdfpageheight\paperheight
\pdfpagewidth\paperwidth

\@ifundefined{textcolor}{}
{%
 \definecolor{BLACK}{gray}{0}
 \definecolor{WHITE}{gray}{1}
 \definecolor{RED}{rgb}{1,0,0}
 \definecolor{GREEN}{rgb}{0,1,0}
 \definecolor{BLUE}{rgb}{0,0,1}
 \definecolor{CYAN}{cmyk}{1,0,0,0}
 \definecolor{MAGENTA}{cmyk}{0,1,0,0}
 \definecolor{YELLOW}{cmyk}{0,0,1,0}
}

\@ifundefined{definecolor}
 {\usepackage{color}}{}
\@ifundefined{definecolor}{\@ifundefined{definecolor}
 {\usepackage{color}}{}
}{}\@ifundefined{definecolor}{\@ifundefined{definecolor}{\@ifundefined{definecolor}
 {\usepackage{color}}{}
}{}}{}\def\1{1\negthickspace{\rm I}}

\newcommand{\PRA}{\mbox{Phys.\ Rev.\ A}}\newcommand{\PRL}{\mbox{Phys.\ Rev.\ Lett.}}

\makeatother

\usepackage{babel}

\makeatother

\usepackage{babel}
\begin{document}

\title{Bipartite entanglement of quantum states in a pair basis}

\author{Marco Roncaglia}

\affiliation{Dipartimento di Fisica del Politecnico, corso Duca degli Abruzzi
24, 10129 Torino, Italy}

\affiliation{INRIM, strada delle Cacce 91, 10135 Torino, Italy}

\author{Arianna Montorsi}

\affiliation{Dipartimento di Fisica del Politecnico, corso Duca degli Abruzzi
24, 10129 Torino, Italy}

\author{Marco Genovese}

\affiliation{INRIM, strada delle Cacce 91, 10135 Torino, Italy}

\date{\today}
\begin{abstract}
The unambiguous detection and quantification of entanglement is a hot topic of scientific research, though it is limited to low dimensions or specific classes of states. Here we identify an additional class of quantum states, for which bipartite entanglement measures can be efficiently computed, providing new rigorous results. 
Such states are written in arbitrary $d\times d$ dimensions, where each basis state in the subsystem A is paired with only one state in B. 
This new class, that we refer to as \textit{pair basis states}, is remarkably relevant in many physical situations, including quantum optics.   
We find that negativity is a necessary and sufficient measure of entanglement for mixtures of states written in the same pair basis. 
We also provide analytical expressions for a tight lower-bound estimation of the entanglement of formation, a central quantity in quantum information. 

\end{abstract}

\pacs{03.67.Mn, 03.65.Ud, 42.65.Lm }

\maketitle
Quantum entanglement, after having been considered for many years a peculiar aspect of quantum mechanics whose interest was limited to 
specialists in foundations in quantum mechanics \cite{Genovese_review05}, has assumed a pervasive role in contemporary science with applications that range from cosmology \cite{1} to biology \cite{2}.  
In particular it attracts a considerable interest being a fundamental resource for quantum technologies  \cite{Zyczkowski_Book,nielsen&chuang2000,AmicoReview08,Genovese_review05,campos06}.
This widespread relevance prompted the need of its unambiguous detection and quantification, a result still largely unachieved \cite{Toth_review09,Facchi07} and recently subject to a large theoretical effort \cite{3}. 
Indeed, while bipartite entanglement in a pure state can be estimated using the von Neumann
entropy as well as other measures, the problem of its
evaluation is still open in the case of a general mixed state. A significant
step forward has been done with the proof that that the bipartite
entanglement in a general mixed state of a system of dimension $2\times2$
is suitably quantified by the concurrence \cite{wootters98}. In $d\times d$
dimensions, with $d\geq3$ \cite{Horodecki96}, a suitable measure
has not yet been found except in the presence of special symmetries,
like in the case of Werner states \cite{werner89}. A computable entanglement monotone 
in arbitrary dimension is the negativity \cite{negativity02}, which
can be evaluated for different physical systems \cite{anfossi05,Calabrese12,sb},
but in general it represents only a sufficient condition for entanglement. 

From the perspective of quantum information, it is helpful to evaluate the entanglement of formation (EOF), a faithful measure 
that quantifies the minimal entanglement resources needed to prepare a given state, in terms of Bell pairs. 
At variance with the negativity, the EOF is typically hard to calculate, as it is obtained from an optimization problem 
that only in some special cases can be solved analytically \cite{wootters98}. 
Sometimes, in higher dimensional system (when exact EOF is impractical to evaluate),  
it is convenient to determine a lower bound to EOF which sets the amount of resources that are \emph{at least} present 
inside the quantum state in consideration. Of course, such a lower bound should be as tight as possible, in order to be useful. 
 
We define the set of pure \emph{pair basis} states of a bipartite system in arbitrary dimension $d\times d$
as the set of states of the form
\begin{equation}
|\Psi\rangle=\sum_{i=1}^{d}c_{i}|\phi_{i}\rangle_{A}\otimes|\chi_{i}\rangle_{B}\:,\label{eq:basic state} 
\end{equation}
where $\{\phi_{i}\}_{i=1}^d$ and $\{\chi_{i}\}_{i=1}^d$ 
are \textit{fixed} orthonormal basis for Hilbert spaces of part $A$ and $B$, respectively. The coefficients $c_i\in \mathbb{C}$ 
satisfy the normalization condition $\sum_{i}|c_{i}|^{2}=1$.
Of course, every pure state can be expressed in the form (\ref{eq:basic state}), through the Schmidt decomposition 
and arbitrary assignment of phase factors to every $c_i$. 
The peculiarity of this set is that all the states belonging to it share the \textit{same} basis, where each element
$\phi_{i}$ in $A$ is paired with only one $\chi_{i}$ in $B$. This represents a subset of the most general bipartite case, 
where quantum states of a given ensemble may have different (pair) basis after Schmidt decomposition.  

In this paper, we address the problem of evaluating the entanglement
of mixtures of states written in a pair basis (\ref{eq:basic state}),
identifying suitable measures. Thus, we estimate rigorous and numerical lower
bounds to the EOF, comparing them to other estimates proposed in the literature \cite{chen05}. 

Pair basis states occur in a variety of physical situations, the most remarkable one being represented by two-mode
Gaussian states in quantum optics \cite{Mandel_Book,Paris_Book05,Genovese_review05,Lopaeva13},
that include twin-beam states, a key element of quantum communication, metrology and sensing.
More interestingly, also non-Gaussian states are included in the family of states of Eq.(1), 
like for example a twin beam plus a dephased component or mixtures of photon subtracted states. 
For these latter cases, that are relevant in many modern experiments of quantum mechanics \cite{gomes09}, 
good measures of entanglement have not been found, so far.   
Moreover, pair basis states are a natural way to explore high-dimensional entanglement, recently observed in spatial 
modes of pairs of down-converted photons \cite{krenn14}.   
In atomic physics, bosonic atoms trapped in
double wells have the same structure (\ref{eq:basic state})
where the conservation of the total number $N$ imposes that each
state $|n\rangle$ of $n$ bosons in one well is paired with the state
$|N-n\rangle$ in the other well. Also for electron models in a lattice
may be interesting to restrict the total Hilbert space of two-sites
to the pair basis $\{\left|\uparrow,\downarrow\right\rangle $, $\left|\downarrow,\uparrow\right\rangle $,
$\left|\downarrow\uparrow,0\right\rangle $, $\left|0,\downarrow\uparrow\right\rangle \}$,
i.e. the sector of zero magnetization and two electrons, thanks to
the presence of special quantum numbers.

\section{Pure states }

Given the orthonormality of both bases $|\phi_{i}\rangle_{A}$ and
$|\chi_{i}\rangle_{B}$, the Schmidt decomposition of states of the
form (\ref{eq:basic state}) is just written in the same basis, but
with the non negative coefficients $\mu_{i}=|c_{i}|$. The Schmidt
coefficients $\mu_{i}$ are the square roots of the eigenvalues of
the reduced density matrix $\rho_{A}=\mathrm{Tr}_{B}\rho$. It follows
that the only factorized states are those with $c_{i}=0$ for
every $i$ except one. The entanglement of a pure state $|\Psi\rangle$
is estimated by the von Neumann entropy $S^{(d)}(\Psi)=-\sum_{i=1}^{d}\mu_{i}^{2}\log\mu_{i}^{2}$.
In the following, we shorten $|\phi_{i}\rangle_{A}\otimes|\chi_{i}\rangle_{B} \equiv |i,i\rangle$.

For estimating the entanglement of pure states, in this work we use
a generalization of the concurrence that we construct in the following
way. First, let us consider the easiest case $d=2$ with the two elements
of the basis $\left|0,0\right\rangle $ and $\left|1,1\right\rangle $.
In the full 4-dimensional space the double spin-flip operation is
performed by $\sigma_{A}^{y}\otimes\sigma_{B}^{y}$, and the concurrence
is given by the well-known formula $C=|\langle\Psi|\sigma_{A}^{y}\otimes\sigma_{B}^{y}|\Psi^{*}\rangle|$, where $\Psi^{*}$ is the complex conjugate of $\Psi$ and $\sigma^{\alpha}$, $\alpha=x,y,z$ are the usual Pauli matrices acting on single qubits.
In our special $d=2$ case for states in a pair basis, the concurrence
becomes $C=|\langle\Psi|\tau^{x}|\Psi^{*}\rangle|=2|c_{1}c_{2}|$,
where the swap is represented by the single Pauli matrix $\tau^{x}$ acting on the pair basis. 

Any extension to the $d$-dimensional case has to take into account the fact that 
for product states, the concurrence has to be zero in every possible two-dimensional subspace. 
Even if the literature proposes different generalizations of the concurrence \cite{audenaert01}, 
we opt to take the sum over every two-dimensional concurrence
\begin{equation}
D(\Psi)=2\sum_{i<j}|c_{i}c_{j}|=(\mathrm{Tr}\sqrt{\rho_{A}})^{2}-1.\label{eq:D1}
\end{equation}
which manifestly vanishes only in the factorized case and likewise it fulfills the additivity property 
whenever the dimension $d$ is built up by the direct product of two-dimensional states. 
In our picture, each term of Eq.(\ref{eq:D1}) considers the state $|j\rangle$ as the spin-flipped of $|i\rangle$. 
A remarkable fact that further justifies the choice of the quantity $D(\Psi)$ is that it turns out to be twice the negativity 
$
\mathcal{N}(\rho)\equiv (\left\Vert \rho^{T_{A}}\right\Vert _{1}-1)/2
$,
where $\rho^{T_{A}}$ stands for the partial transpose with respect
to subsystem $A$ and $\left\Vert G\right\Vert _{1}=\mathrm{Tr}\sqrt{GG^{\dagger}}$
is the trace norm. For proving this statement we write the entries
of the density matrix of the state (\ref{eq:basic state}) in the
full Hilbert space
$
\left\langle i,j\right|\rho\left|i',j'\right\rangle =\delta_{ij}\delta_{i'j'}c_{i}c_{i'}^{*}.
$
As a matter of fact, the operation of partial transposition introduces
non vanishing matrix elements outside the set of pair basis states, obtaining 
$
\left\langle i,j\right|\rho^{T_{A}}\left|i',j'\right\rangle =\delta_{ij'}\delta_{i'j}c_{i}c_{i'}^{*}.
$
Besides the diagonal part of $\rho$ that is left unchanged,
the matrix $\rho^{T_{A}}$ displays $2\times2$ blocks for every pair $i<j$ in the subspace formed by the two basis vectors
$\left|i,j\right\rangle $ and $\left|j,i\right\rangle $. The eigenvalues
of such blocks turn out to be pairs of opposite numbers $\pm|c_{i}c_{j}|$,
signaling the presence of entanglement due to a negative eigenvalue.
So, the negativity for pure pair states amounts to 
$\mathcal{N}(\Psi)=\sum_{i<j}|c_{i}c_{j}|$, which is a necessary and sufficient measure of entanglement, 
being zero only for factorizable states.    

\section{Mixed states}

Mixtures of pure states in the same pair basis, define a large nontrivial subset of the full Hilbert space, 
described by density matrices depending on $d^2-1$ independent parameters.   
A fundamental property of the negativity is that it represents an
entanglement monotone under local operations and classical communication
(LOCC) for every mixed state in arbitrary dimension \cite{Zyczkowski_Book,negativity02}.
Specifically, $\mathcal{N}(P(\rho))\leq\mathcal{N}(\rho)$ for an arbitrary LOCC $P(\rho)$. 
Moreover, given that a mixed state is a convex combination of pure states $\rho_i$, 
the negativity is in general a convex function,
i.e. $\mathcal{N}\left(\sum_{i}p_{i}\rho_{i}\right)\leq\sum_{i}p_{i}\mathcal{N}(\rho_{i})$,
with weights obeying to $\sum_{i}p_{i}=1$ and $p_{i}\geq0$, $\forall i$. 

For a mixed pair state $\rho_{ij}$, a convex combination of pure states in a pair basis (\ref{eq:basic state}), 
the negativity becomes 
\begin{equation}
\mathcal{N}(\rho)=\sum_{i<j}|\rho_{ij}|\label{eq:Neg_our}
\end{equation}
which turns out to be a good measure of entanglement for our class of states, 
since $\mathcal{N}$ is convex and vanishes only in absence of off-diagonal terms of $\rho_{ij}$, i.e. for factorizable
states. In other words, this proves that a \emph{necessary} and \emph{sufficient} condition
for having entanglement is the non vanishing of negativity, a property
which is not valid for general states. In addition, the monotonicity
of $\mathcal{N}(\rho)$ introduces a ordering in terms of entanglement
content. Despite its simplicity, Eq.(\ref{eq:Neg_our}) constitutes
an important result, which may reveal of great utility in the evaluation
of entanglement in several systems expressable in a pair basis. Along
the same line, one can compute the logarithmic negativity $E_{\mathcal{N}}(\rho)\equiv\log\left\Vert \rho^{T_{A}}\right\Vert _{1}=\log(1+2\sum_{i<j}|\rho_{ij}|)$,
which bounds the distillable entanglement of $\rho$ \cite{negativity02}.

\section{Entanglement of formation}

The EOF $E_{f}(\rho)$ is in general defined as the convex roof 
\begin{equation}
E_{f}(\rho)=\mathrm{\min_{\left\{ p_{k},\psi_{k}\right\} }}\sum_{k=1}^{d}p_{k}S^{(d)}(\psi_{k}).\label{eq:EOF}
\end{equation}
that gives the minimum average entropy over all possible decompositions
of $\rho=\sum_{k}p_{k}|\psi_{k}\rangle\langle\psi_{k}|$ into pure
states $|\psi_{k}\rangle$, $k=1,\dots d$, with weights $p_{k}$.
The calculation of (\ref{eq:EOF}) for general states is notoriously
a formidable task. However, for our class of states we are able to
establish some tight lower bounds of $E_{f}$ of evident use in a
variety of applications. 

In our case, the task is somewhat facilitated by the crucial property
that the element $|\psi_{k}\rangle$ of every decomposition
must also be restricted to the same pair basis as $\rho$. In fact,
the diagonal elements $\langle ij|\rho|ij\rangle=0$ for $i\neq j$,
can only result from a convex combination of zero diagonal elements
$\langle ij|\rho_{k}|ij\rangle=0$ for any $k$, since they cannot be negative by
definition, as well as the weigths $p_{k}$. As a consequence, even
the off-diagonal elements of $\rho_{k}$ that lie outside the pair
basis are zero. 

The entanglement of any pure state in Eq. (\ref{eq:basic state})
is determined by $d-1$ parameters, the Schmidt weights $\mu_{j}=|c_j|$,
plus the normalization condition. As known, these coefficients cannot
be inferred by the partial trace in the mixed case. Instead, we want
to relate them to the off-diagonal entries of $\rho_{ij}$. After
relabeling the states such that $\mu_{1}\geq\mu_{2}\geq\dots\geq\mu_{d}$
(hence $\Gamma_{1}\geq\Gamma_{2}\geq\dots\geq\Gamma_{d}$), it holds that 
\begin{align}
\mu_{i}^{2}= & \frac{1}{2}\left(1-\epsilon_{i}\sqrt{1-4\Gamma_{i}^{2}}\right)\label{eq:mu_i alt}
\end{align}
where 
\begin{equation}
\Gamma_{i}^{2}=\mu_{i}^{2}(1-\mu_{i}^{2})=\sum_{j\neq i}\mu_{i}^{2}\mu_{j}^{2}=\sum_{j\neq i}|\rho_{ij}|^{2}\label{eq:Gamma_i}
\end{equation}
and all $\epsilon_{i}=1$ except $\epsilon_{1}=-1$ if $|\mu_{1}|^{2}>1/2$. 
Of course, for pure states the quantities
$|\rho_{ij}|^{2}$ are overdetermined, so there are many ways to take
$d-1$ of them which are independent. One way to avoid such an overdetermination
is to consider only the first row of the density matrix, obtaining
$\mu_{j}^{2}=|\rho_{1j}|^{2}/\mu_{1}^{2},\; j=2,\dots,d$, and $\mu_{1}^{2}$
as in Eq.(\ref{eq:mu_i alt}). Now, we are
able to find a lower bound to the EOF, by means of the following:

\paragraph{Theorem }

For every pair state described by a density matrix $\rho$ in the relabeled
basis with $\Gamma_{1}\geq\Gamma_{2}\geq\dots\geq\Gamma_{d}$, it holds
\begin{equation}
E_{f}(\rho)\geq F(\mathbf{x})\equiv-\sum_{i=1}^{d}\alpha_{i}^{2}(\mathbf{x})\log\alpha_{i}^{2}(\mathbf{x})\label{eq:F(x)}
\end{equation}
where
\begin{equation}
\alpha_{1}^{2}=\frac{1}{2}\left(1+\sqrt{1-4|\mathbf{x}|^{2}}\right);\quad\alpha_{i}^{2}=\frac{|x_{i}|^{2}}{\alpha_{1}^{2}},\: i=2,\dots,d\label{eq:alphas}
\end{equation}
and the components of the vector $\mathbf{x}\equiv\{\rho_{12},\rho_{13},\dots,\rho_{1d}\}$
are the $d-1$ off-diagonal elements in the first row of $\rho$,
which act as independent parameters. 

\paragraph{Proof}

Let us assume that there exist an optimal decomposition of 
$\rho=\sum_{k}p_{k}|\psi_{k}\rangle\langle\psi_{k}|$
formed by an ensemble of pure states $\left\{ p_{k},|\psi_{k}\rangle\right\} $, 
where each $\psi_{k}$ must belong to the same pair basis set, as discussed above. 
To any state $\psi_{k}$ of the decomposition, we associate a set of vectors $\mathbf{x}^k$ 
and exact Schmidt weights $\{\mu_{i}^{2}(\psi_{k})\}_{i=1}^d$, as given by  Eqs.(\ref{eq:mu_i alt}) and (\ref{eq:Gamma_i}). 
The off-diagonal elements of $\rho$ contained in $\mathbf{x}$ are given by the convex sum
$\mathbf{x}=\sum_{k}p_{k}\mathbf{x}^{k}$. It follows that the EOF is lower bounded
by $F(\mathbf{x})$ because
\begin{align}
E_{f}(\rho) & =-\sum_{k}p_{k}\left(\sum_{i=1}^{d}\mu_{i}^{2}(\psi_{k})\log\mu_{i}^{2}(\psi_{k})\right)\nonumber \\
 & \geq-\sum_{k}p_{k}\left(\sum_{i=1}^{d}\alpha_{i}^{2}(\mathbf{x}^{k})\log\alpha_{i}^{2}(\mathbf{x}^{k})\right)\nonumber \\
 & =\sum_{k}p_{k}F(\mathbf{x}^{k})\geq F(\mathbf{x}).\label{eq:LB_EOF}
\end{align}
The first inequality has been obtained by observing that $\forall z\in[0,1]$,
we get $\frac{1}{2}\left(1+\sqrt{1-z}\right)\log\left[\frac{1}{2}\left(1+\sqrt{1-z}\right)\right]\geq\frac{1}{2}\left(1-\sqrt{1-z}\right)\log\left[\frac{1}{2}\left(1-\sqrt{1-z}\right)\right]$, that gives a lower bound for the first term in the sum in (\ref{eq:LB_EOF}), and eliminates the problem of determining the sign $\epsilon_{1}$. 
This latter inequality is also crucial to compensate the increase of the second term in the sum, when $e^{-1}\leq \mu_{1}^{2}\leq\mu_{2}^{2}\leq1/2$. 
The second inequality in (\ref{eq:LB_EOF}) holds thanks to the convexity of $F(\mathbf{x})$
over its domain $\{\mathbf{x}:\:|\mathbf{x}|^{2}\in[0,\frac{1}{4}]$\}.
After observing that $F$ depends only on the moduli of the components
of $\mathbf{x}=\{x_i\}_{i=2}^d$, we get 
\begin{align*}
\sum_{k}p_{k}F(\mathbf{x}^{k}) & =\sum_{k}p_{k}F(\{|x_{i}^{k}|\})\geq F(\{\sum_{k}p_{k}|x_{i}^{k}|\})\\
 & \geq F(\{|\sum_{k}p_{k}x_{i}^{k}|\})=F(\mathbf{x}),
\end{align*}
where the first inequality comes from the convexity of $F(\mathbf{v})$
as a function of a real positive vector $\mathbf{v}$, such that $|\mathbf{v}|^{2}\in[0,\frac{1}{4}]$,
while the second inequality is a consequence of the triangular inequality 
$|z+w|\leq|z|+|w|$, with $z,w\in\mathbb{C}$ and the monotonicity
of $F(\mathbf{v})$ with respect to any of its components. The latter
property can be directly proven by calculating $\nabla[F(\mathbf{x})]$
and showing that all its components are non negative for $|\mathbf{v}|^{2}\in[0,\frac{1}{4}]$ (see Appendix \ref{convF}). 
Proving that $F(\mathbf{v})$ is convex for any $d$ by checking directly the positiveness of its Hessian $\mathcal{H}[F(\mathbf{v})]$ 
can be a hard task. By means of a mathematical stratagem we have found a way to write Eq.(\ref{eq:F(x)}) as 
a sum of convex functions $F(\mathbf{v})=\sum_k F_k(\mathbf{v})$, 
where each $F_k$ explicitly depends only on the two variables $\vert\mathbf{v}\vert$ and $v_k$, 
whose convexity is analytically proven through ordinary calculus methods (see Appendix \ref{convF}). 
This shows that $F(\mathbf{v})$ is indeed convex in every dimension. $\square$ 

\begin{figure}
\includegraphics[scale=0.215]{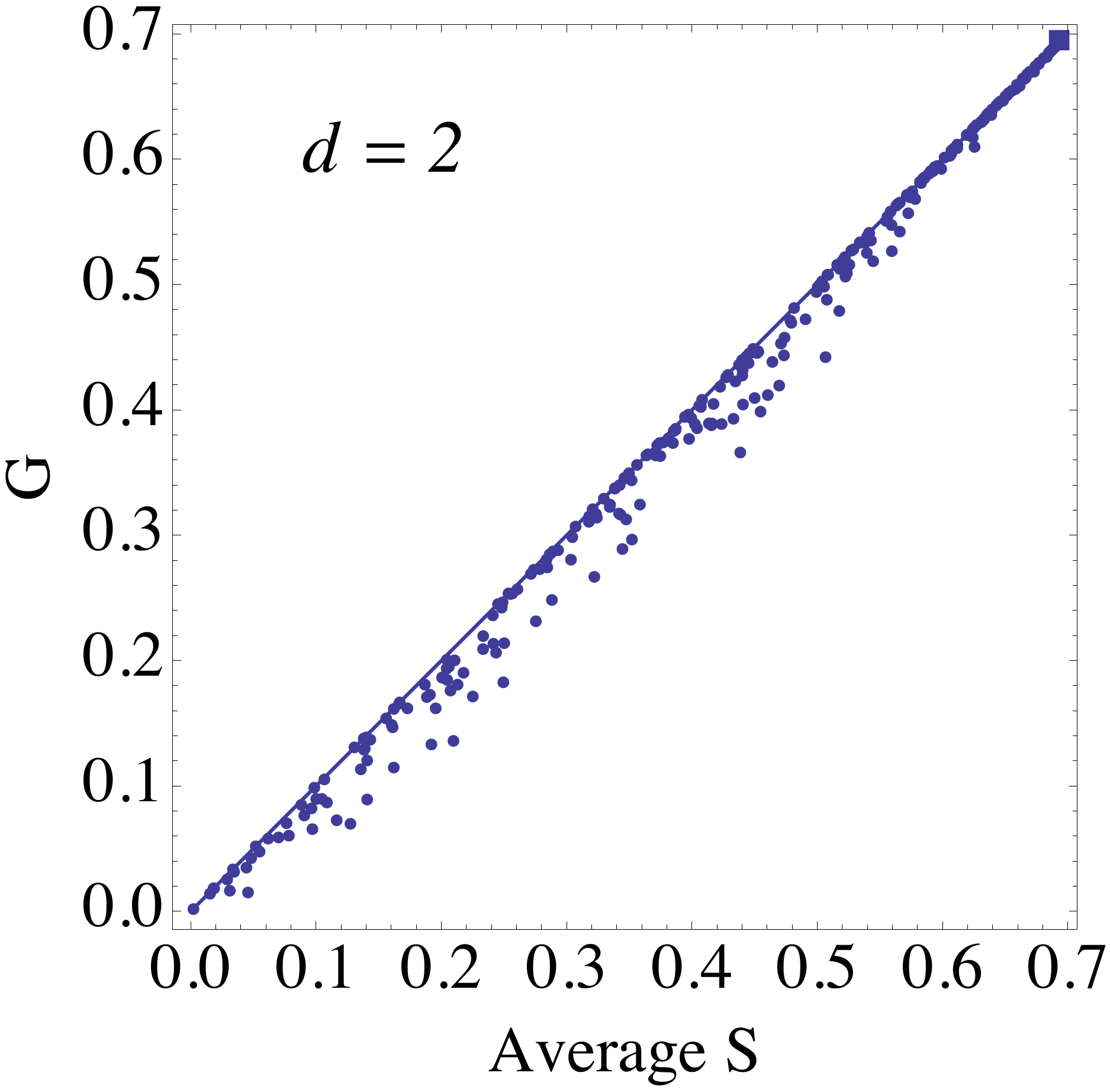}~~\includegraphics[scale=0.21]{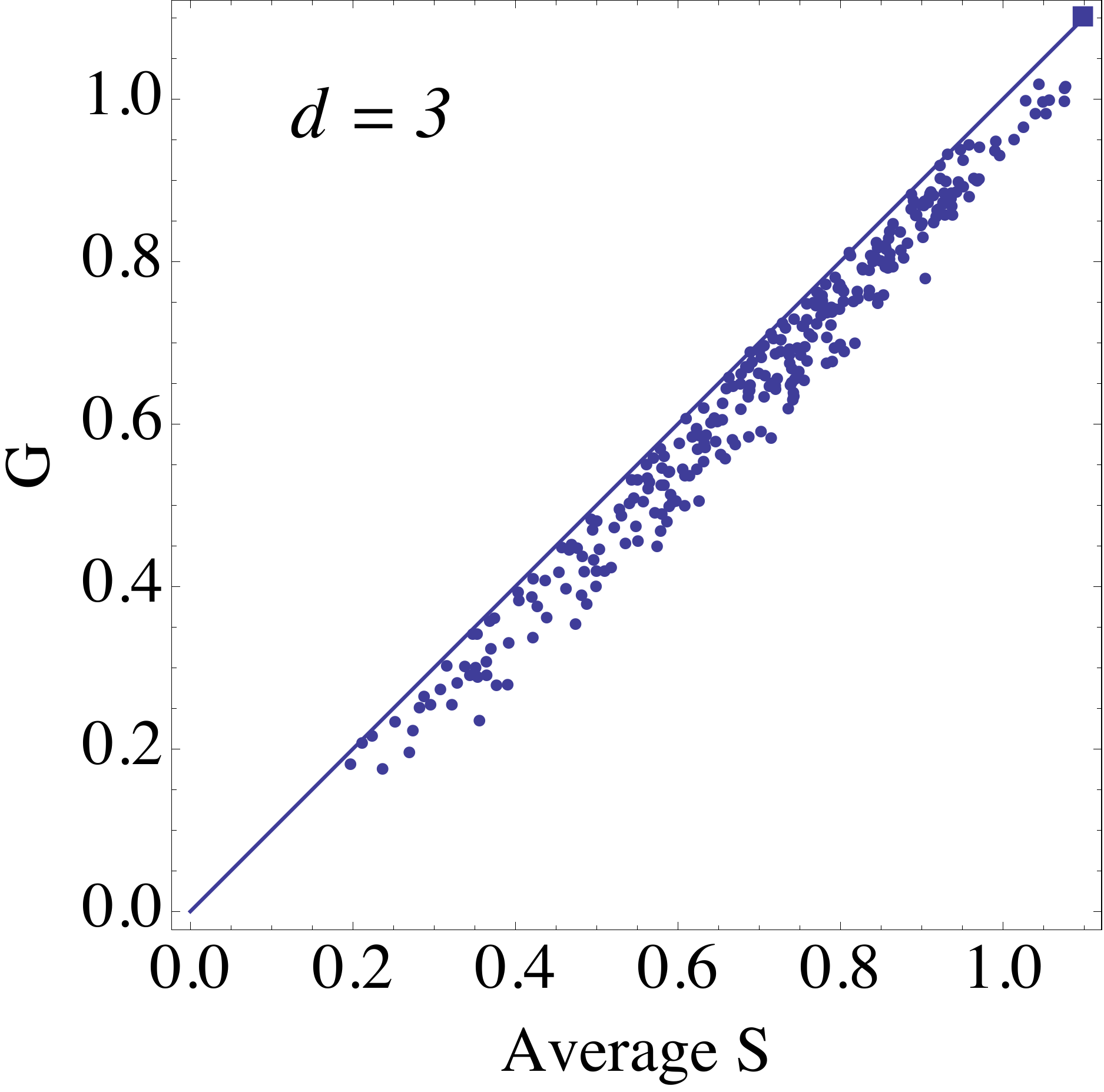}
\medskip{}
\includegraphics[scale=0.215]{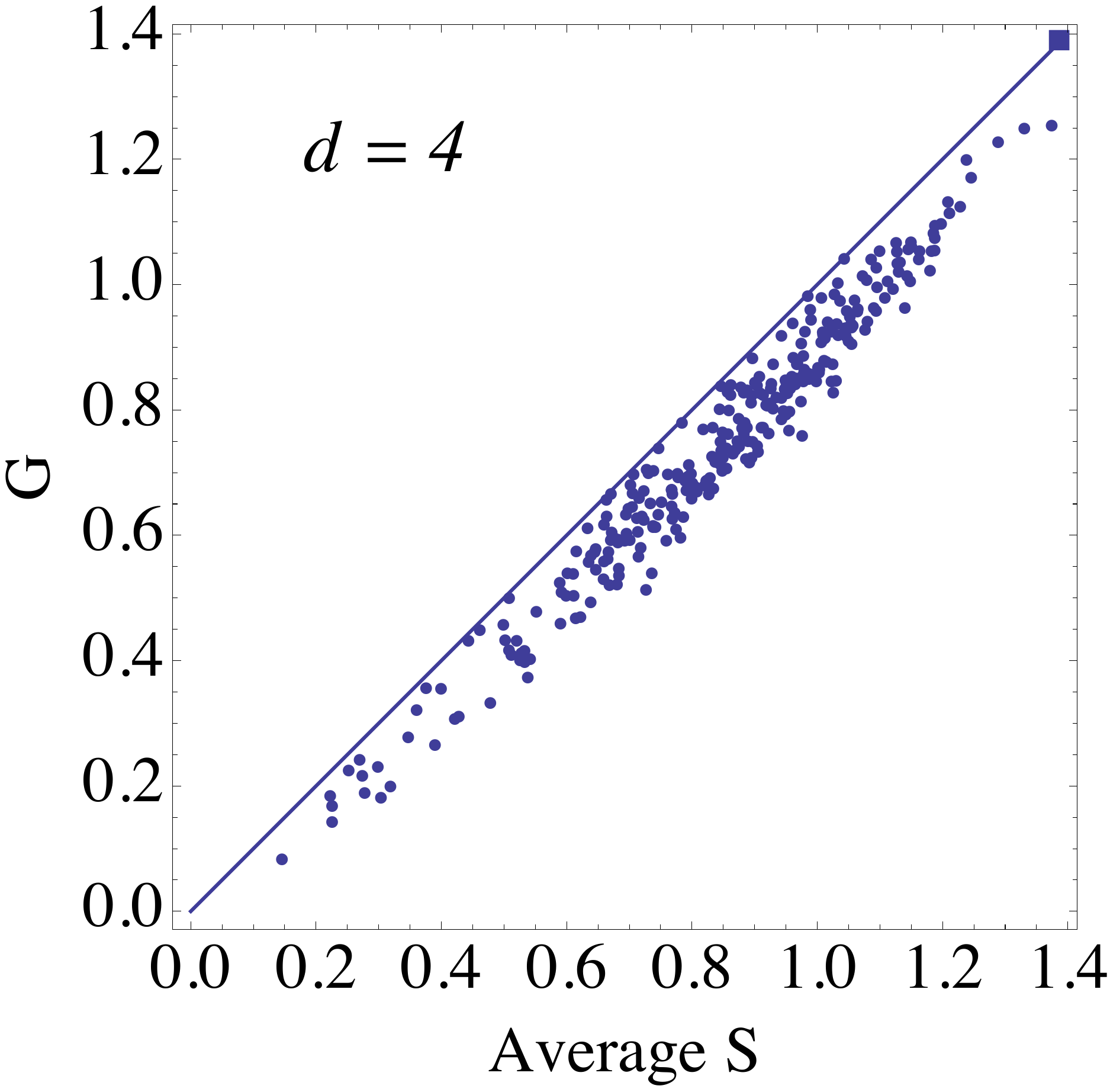}~~\includegraphics[scale=0.21]{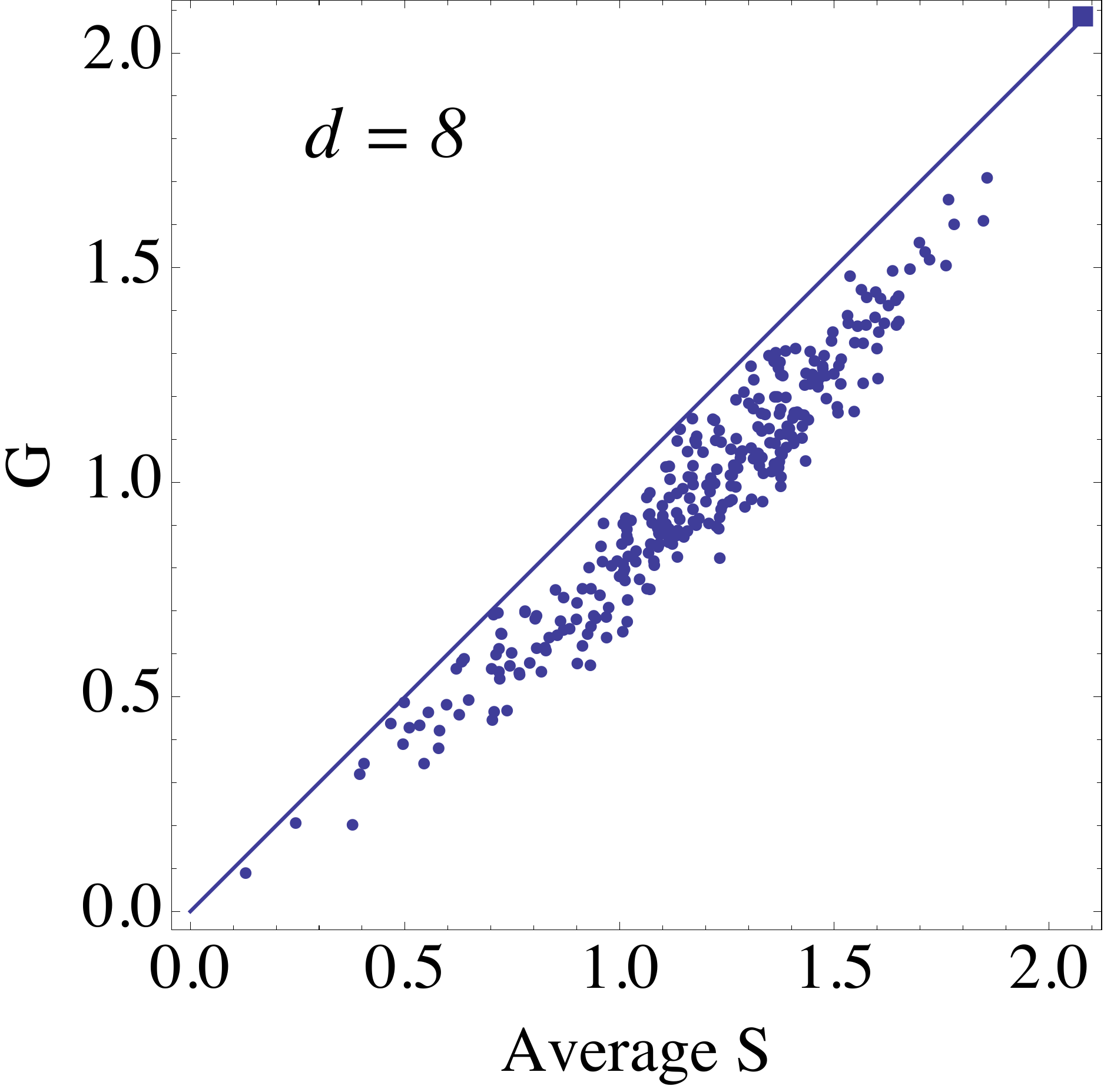}
\caption{(Color online) The average entropy (horizontal axis) is plotted against the
function $G$ (vertical axis) for several density matrices $\rho$
in various dimensions $d$. Each point is calculated from a set of
randomly generated pure states, combined with random weights (see text). The
fact that all the points lie below the bisector line, constitutes
a stochastic demonstration that $G$ is a lower bound to the EOF.\label{fig:G random}}
\end{figure}

A step further in the seek of lower bounds for the EOF can be made
by defining the function $G(\mathbf{x}_{1},\dots,\mathbf{x}_{d})$
in a similar way as $F(\mathbf{x})$ in Eqs. (\ref{eq:F(x)}) and
(\ref{eq:alphas}), but with 
\begin{align*}
\alpha_{1}^{2} & =(1+\sqrt{1-4|\mathbf{x}_{1}|^{2}})/2,\\
\alpha_{i}^{2} & =(1-\sqrt{1-4|\mathbf{x}_{i}|^{2}})/2,\quad i=2,\dots,d
\end{align*}
where $\mathbf{x}_{i}\equiv\{\rho_{ii'},\: i'\neq i\}$. In this
new definition the number of independent variables is increased to
$d(d-1)/2$. Unfortunately, the function $G$ is not convex over all
its domain, a property that would be a sufficient condition for proving
its validity as a lower bound for EOF, like in the previous theorem.
However, strictly speaking we need only the weaker property of convexity
with respect to the set of pure states (as opposed to the overall
set of density matrices). This latter feature is indeed displayed
by $G$, though it seems to be rather hard to prove it analytically.
Instead, we provide a stochastic demonstration by generating several
samples of $\rho$ through random mixtures of pure states $\{ p_{k},\psi_{k}\}$ 
with uniformly distributed $p_k$ and coefficients of $\psi_{k}$. 
We have numerically checked that the average entropy $\sum_{k}p_{k}S^{(d)}(\psi_{k})$ 
of every sampled decomposition is always larger than the lower bound $G$, calculated directly 
from the matrix elements of $\rho$. Without
loss of generality, we have generated pure states of the form (\ref{eq:basic state})
with real positive coefficients, i.e. pure state decompositions which
are closer to the optimal one (that we do not know) thank to the triangular
inequality and because the entropy depends only on the absolute
value of the entries of $\rho$. The results of the numerical simulations
are collected in Fig.\ref{fig:G random}. 

Finally, we consider the lower bound introduced in Ref. \cite{chen05},   
that depends on $\rho$ only through one variable: 
the maximum between the partial transpose and the realignment, for general states.
Restricting to pair states, this single parameter reduces to the negativity. 
In the set of pure states, there is no
unique correspondence between negativity and entropy for $d>2$, as  
there can be states with the same negativity but different entropy. However,
one may introduce a convex function $s(\mathcal{N})$, that for any
$\mathcal{N}$ is not larger than the minimum entropy 
$S_{\mathrm{min}}^{(d)}(\mathcal{N})=\mathrm{\min_{\{\psi_{k},\mathcal{N}\}}}S^{(d)}(\psi_{k})$
in the manifold of all pure states $|\psi_{k}\rangle$ with a given negativity
$\mathcal{N}$. This optimization problem has been solved in 
\cite{terhal00}, with solution 
\[
s(\mathcal{N})=\begin{cases}
H_{2}(\gamma)+(1-\gamma)\log(d-1), & \mathcal{N}\in\left[0,\frac{3}{2}-\frac{2}{d}\right]\\
\frac{2\mathcal{N}+1-d}{d-2}\log(d-1)+\log d, & \mathcal{N}\in\left[\frac{3}{2}-\frac{2}{d},\frac{d-1}{2}\right]
\end{cases}
\]
where 
$
\gamma(\mathcal{N})=\frac{1}{d^{2}}[\sqrt{2\mathcal{N}+1}+\sqrt{(d-1)(d-2\mathcal{N}-1)}]^{2}.
$
By assuming first to know the optimal decomposition $\left\{ p_{k},|\psi_{k}\rangle\right\} $
that gives the minimum in Eq.(\ref{eq:EOF}), one can apply the inequalities
\begin{align}
E_{f}(\rho) & \geq\sum_{k}p_{k}\mathrm{\min_{\left\{ p_{k},\psi_{k}\right\} }}S^{(d)}(\psi_{k})\geq\sum_{k}p_{k}s(\mathcal{N}(\psi_{k}))\nonumber \\
 & \geq s\left(\sum_{k}p_{k}\mathcal{N}(\psi_{k})\right)\geq s(\mathcal{N}(\rho)),\label{eq:EOF ineq1}
\end{align}
thanks to the convexity of both $s$ and $\mathcal{N}$. It is clear
that the function $s(\mathcal{N}(\rho))$ sets a lower bound to the
EOF of $\rho$. The advantage of introducing the function $s$ is
to establish a 1-1 relationship between negativity and EOF, like in
the two-qubit case. On the one hand, this lower bound to EOF is exact
for isotropic states \cite{terhal00} and in our case works very well
for states where the off-diagonal terms assume very similar values.
On the other hand, $F$ and $G$ set better lower bounds for states
where few $\Gamma_{j}$'s dominate over the others. In particular,
in large dimensions when $d\gg\mathcal{N}$, the leading term 
$s(\mathcal{N})\approx(2\mathcal{N}-1)d^{-1}\log d$ goes to zero. 
In fact, in infinite dimension $s(\mathcal{N})$ fails to give a reliable lower bound for every finite $\mathcal{N}$, while 
$F$ and $G$ still give a good estimation of EOF, e.g. in the case of two-mode squeezed states (see Appendix \ref{TWB}).

\begin{figure}
\includegraphics[scale=0.215]{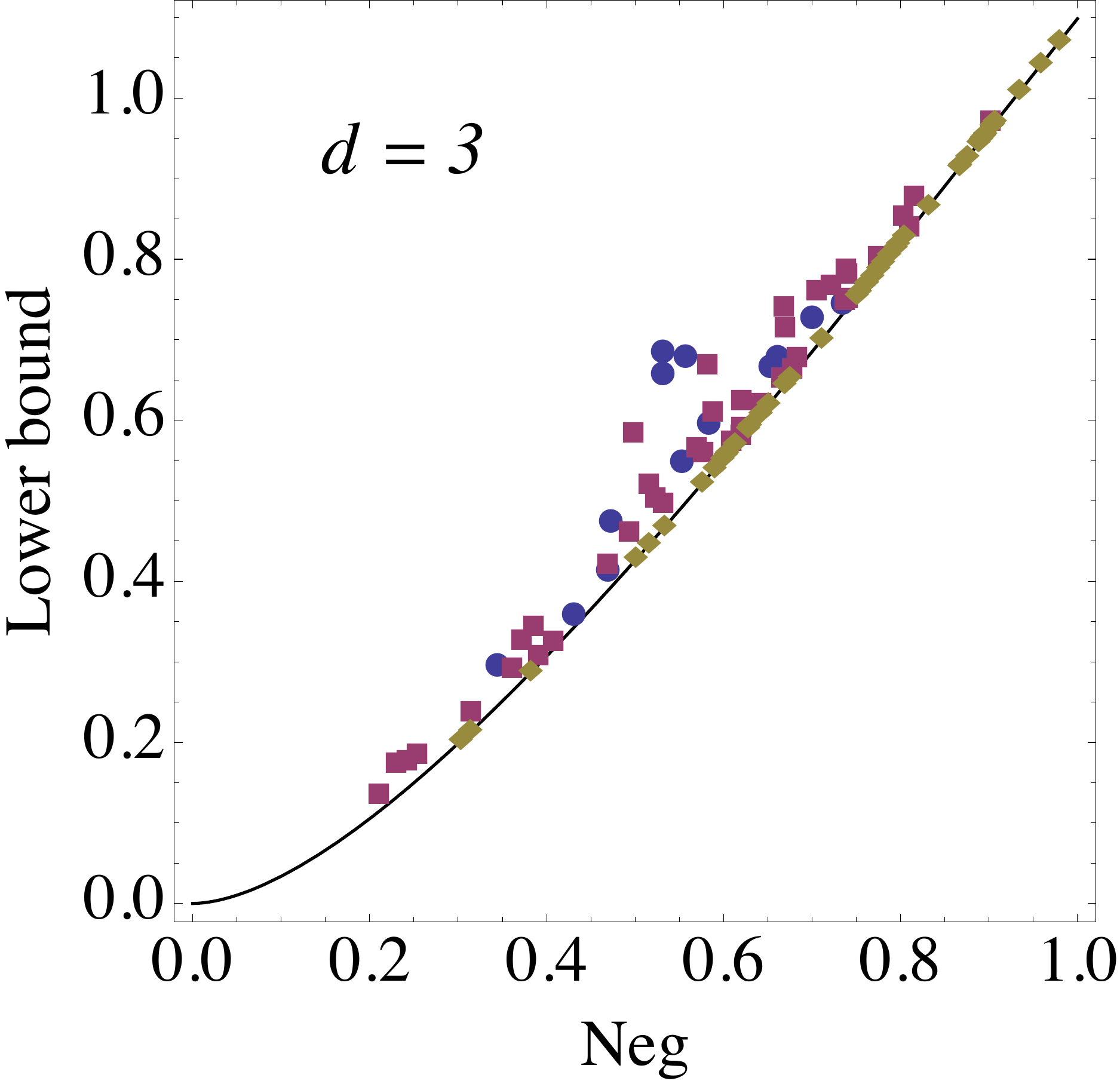}~~\includegraphics[scale=0.21]{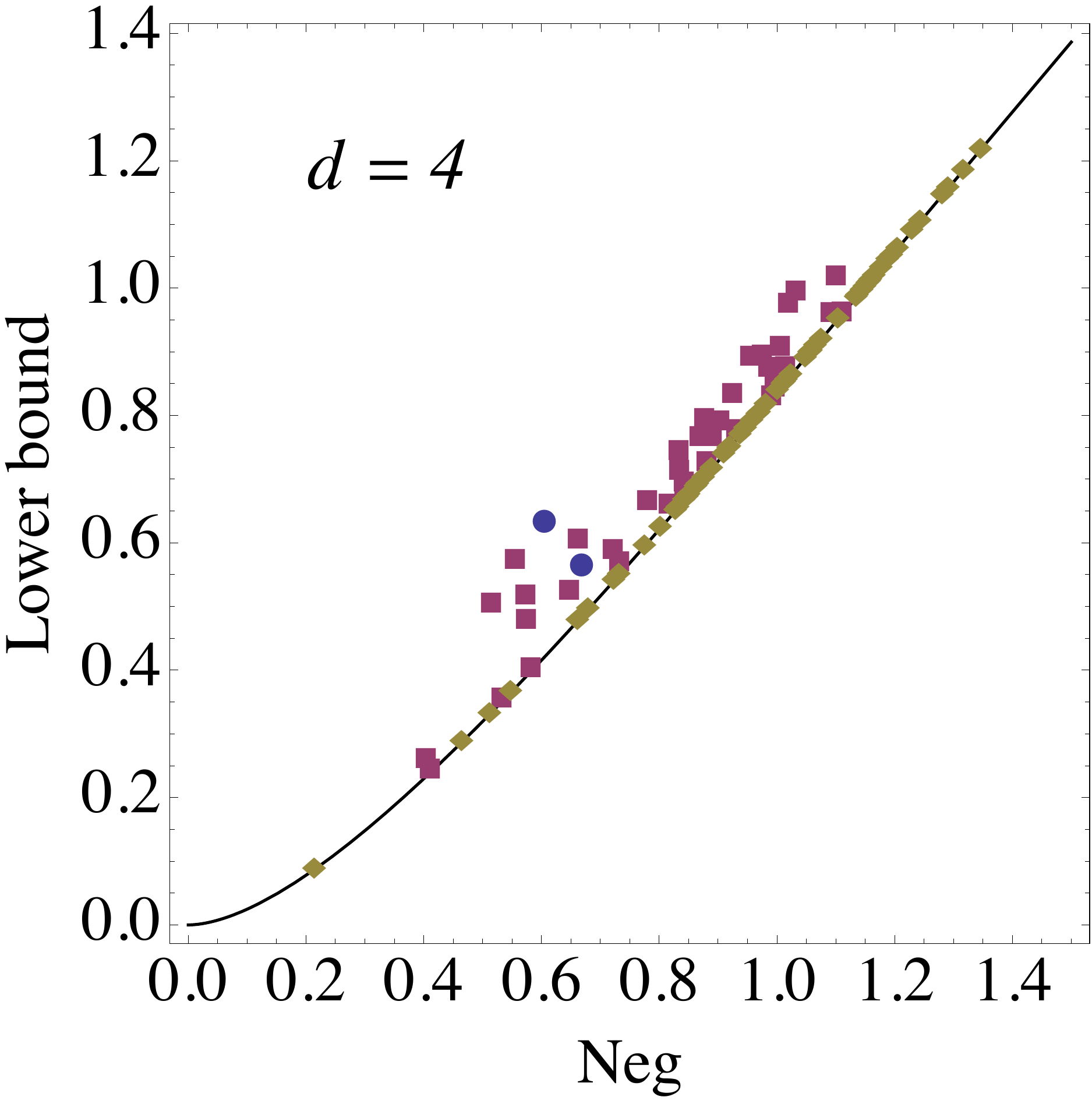}
\medskip{}
\includegraphics[scale=0.215]{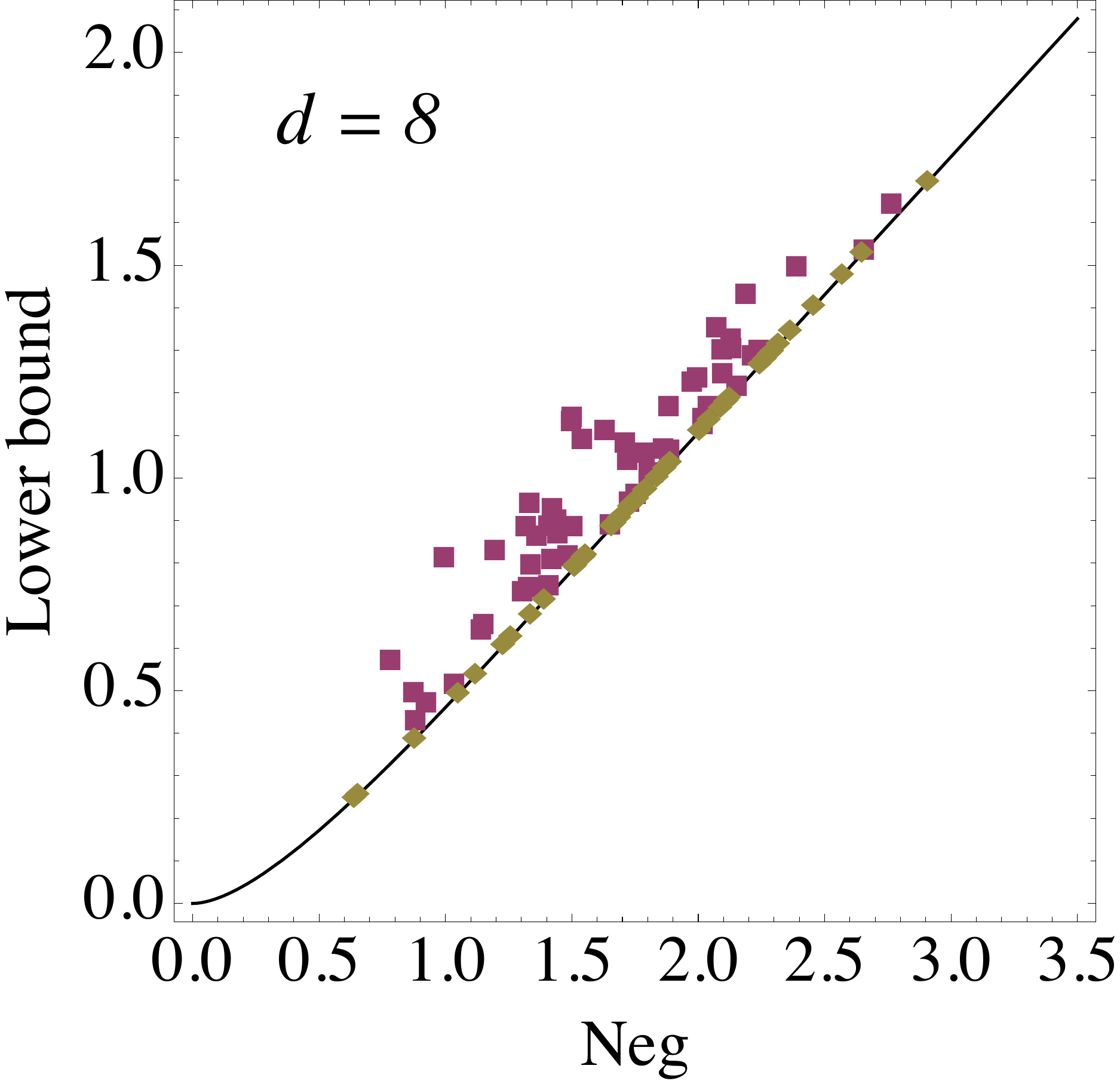}~~\includegraphics[scale=0.21]{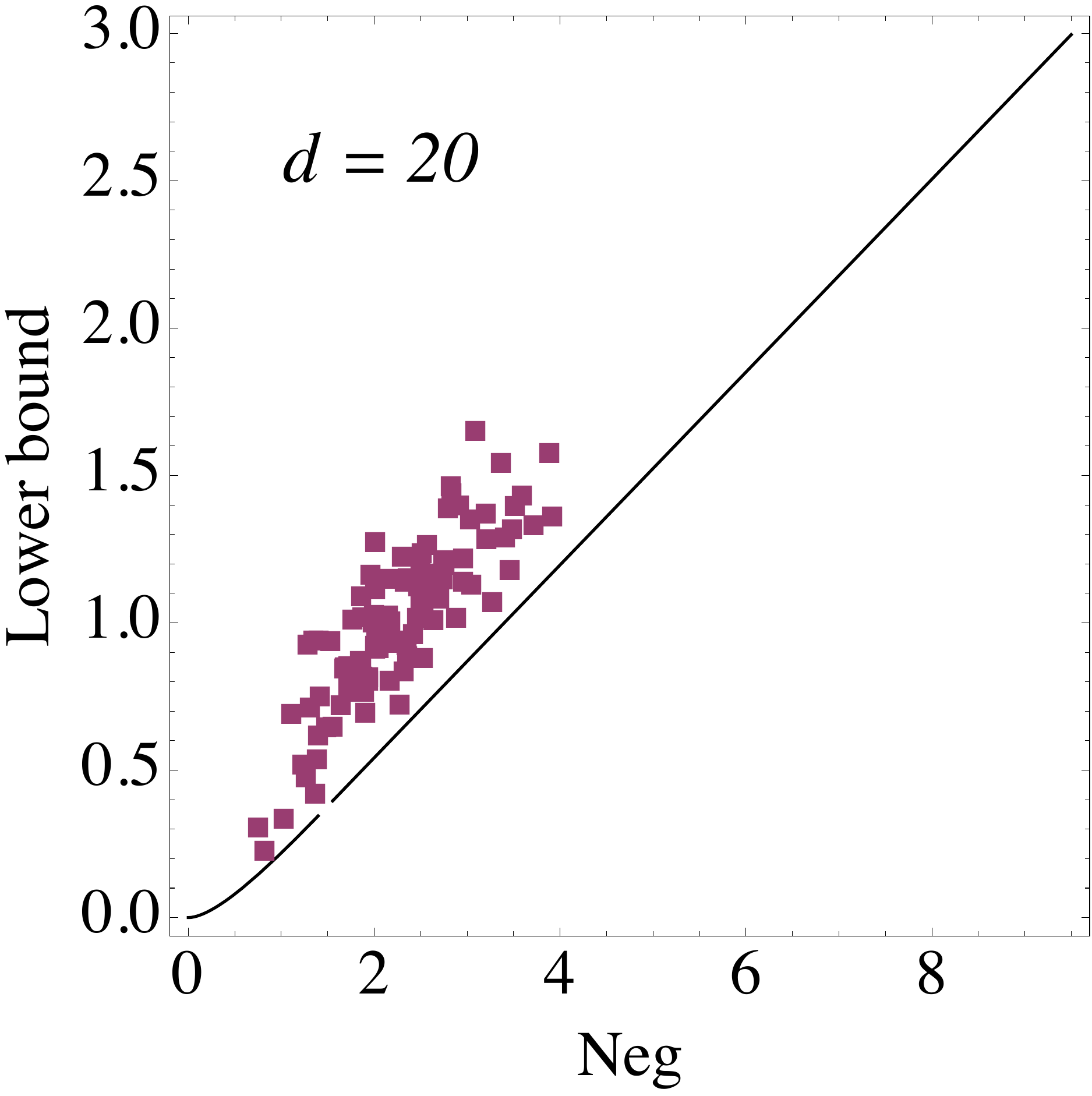}
\caption{(Color online) The maximum between the lower bounds to the EOF in various dimensions
$d$. Circles (blue), squares (red) and rhombs (yellow) represent
$F(\rho)$, $G(\rho)$ and $s(\rho)$, respectively. The sample states
$\rho$ are arranged according to their negativity $\mathcal{N}$
(horizontal axis). For reference, we have plotted the curve $s(\mathcal{N})$.
\label{fig:LB compared}}
\end{figure}

The best estimation of the EOF for an arbitrary state $\rho$ is given by 
$\max\{F(\rho),G(\rho),s(\rho)\}$. In Fig.\ref{fig:LB compared} we show
a comparative plot of the three lower bounds $F,$ $G$, and $s$
for some randomly generated states (with the sampling method used for Fig.\ref{fig:G random}), ordered according to their negativity. 
As expected from the previous analysis, $s$ is a good estimation of EOF for some instances in low dimension and close
to the maximally entangled state. However, as the dimension is increased $s$ becomes useless, while $F$ and $G$ 
give sizable estimates. For high dimensions, we observe that $G$ is slightly greater than $F$, so it tends to dominate, as shown 
in Fig.\ref{fig:LB compared} for $d=20$, where the square symbols often overtake the circles.   
More detailed applications of entanglement estimation of states in pair basis will be considered in forthcoming works. 

\section*{Conclusions }

In this article we have significantly extended the family of states for which the
negativity is a necessary and sufficient condition for entanglement by
considering mixtures of pure states written in the paired form (\ref{eq:basic state}), which are relevant in several physical situations.
We have also found new lower bounds improving the estimation of the
EOF with respect to other quantities known in the literature. 
We believe that our scheme for determining the
functions $F$ and $G$ may be extended to arbitrary states, 
shining some light toward the identification of a general efficient entanglement measure.

\section*{Acknowledgements}

We acknowledge EU-ERC project no. 267915 (OPTINF), the Compagnia di San Paolo and NATO grant SFP 984397  for partial support. 

\appendix
\section{Convexity of the function $F$}\label{convF}

The entropy function $F$, defined in Eq.(\ref{eq:F(x)}),  
\begin{equation}
F(\mathbf{x})\equiv-\sum_{i=1}^{d}\alpha_{i}(\mathbf{x})\log\alpha_{i}(\mathbf{x})
\end{equation}
is notoriously a concave function as a function of the $\alpha$'s.
However, here we want to proof that $F$ as a function of the $(d-1)$-dimensional
vector $\mathbf{x}\equiv\{x{}_{i},\: i=1,\dots,d-1\}$ is instead
convex over the domain $\mathcal{D}=\{x{}_{i}\geq0,|\mathbf{x}|^{2}\leq1/4\}$,
with the choice 
\begin{equation}
\alpha_{1}=\frac{1}{2}\left(1+\sqrt{1-4r^{2}}\right);\quad\alpha_{i}=\frac{x_{i-1}^{2}}{\alpha_{1}},\: i=2,\dots,d\label{eq:alphas1}
\end{equation}
where we have renamed $r=\sqrt{|\mathbf{x}|^{2}}$. 

Prooving the convexity of $F$ in every dimension $d$ through a ``brute
force'' demonstration of the positiveness of its Hessian, can be
a hard task. Here instead we proceed by presenting a detailed calculation after 
splitting the function in a sum of convex terms.  

For convenience, we rewrite $F(\mathbf{x})$ as
\begin{eqnarray}
F & = & -\alpha_{1}\log\alpha_{1}-\sum_{i=1}^{d-1}\frac{x_{i}^{2}}{\alpha_{1}}\log\frac{x_{i}^{2}}{\alpha_{1}}\nonumber \\
&=& (1-2\alpha_{1})\log\alpha_{1}-\frac{1}{\alpha_{1}}\left(\sum_{i=1}^{d-1}x_{i}^{2}\log\frac{x_{i}^{2}}{r^{2}}-r^{2}\log r^{2}\right)\nonumber \\
 & = & \sum_{k=1}^{d-1}\frac{x_{k}^{2}}{r^{2}}\left[-\alpha_{1}\log\alpha_{1}-(1-\alpha_{1})\log(1-\alpha_{1})\phantom{\frac{0}{0}}\right.\nonumber \\
 & & \left. -(1-\alpha_{1})\log\frac{x_{k}^{2}}{r^{2}}\right]\equiv\sum_{k=1}^{d-1}F_{k}(\mathbf{x})\label{eq:F alt}
\end{eqnarray}
where we have used the identity $r^{2}=\alpha_{1}(1-\alpha_{1})$.
We have rewritten $F$ in this way because it is feasible to show
that $F_{k}(\mathbf{x})$ is convex in $\mathcal{D}$ for every $k=1,\dots,d-1$.
Since the sum of convex functions is convex, this will prove our statement. 

\paragraph{Lemma }

The function 
\[
F_{k}(\mathbf{x})=\frac{x_{k}^{2}}{r^{2}}\left[H_{C}(r)-f(r)\log\frac{x_{k}^{2}}{r^{2}}\right]
\]
where 
\begin{eqnarray*}
H_{C}(r) & = & -\frac{1}{2}\left(1+\sqrt{1-4r^{2}}\right)\log\left[\frac{1}{2}\left(1+\sqrt{1-4r^{2}}\right)\right]\\
 &  & -\frac{1}{2}\left(1-\sqrt{1-4r^{2}}\right)\log\left[\frac{1}{2}\left(1-\sqrt{1-4r^{2}}\right)\right]\\
f(r) & = & \frac{1}{2}\left(1-\sqrt{1-4r^{2}}\right)
\end{eqnarray*}
is convex in the domain \emph{$\mathcal{D}$} and it is an increasing
function with respect to any component of $\mathbf{x}$. (Notice that
$H_{C}$ has just the same form as the entropy for a pair of qubits
as a function of the concurence)

\paragraph{Proof }

We use the fact that $F_{k}(\mathbf{x})$ depends explicitly only
on the two quantities $g_{1}(\mathbf{x})=|\mathbf{x}|$ and $g_{2}(\mathbf{x})=x_{k}/r$,
thus 
\begin{eqnarray*}
F_{k}(\mathbf{x})&=&F_{k}(g_{1}(\mathbf{x}),g_{2}(\mathbf{x}))\\
&\equiv & g_{2}^{2}(\mathbf{x})\left[H_{C}(g_{1}(\mathbf{x}))-f(g_{1}(\mathbf{x}))\log g_{2}^{2}(\mathbf{x})\right]
\end{eqnarray*}
whose gradient and Hessian functions in terms of $\mathbf{g}=(g_{1},g_{2})$
are expressed as 
\begin{widetext}
\[
\nabla_{g}[F_{k}(\mathbf{g})]=\left(G_{10},G_{01}\right)=\left(g_{2}^{2}\left(H_{C}^{'}-f^{'}\log g_{2}^{2}\right),2g_{2}\left[H_{C}-f\left(1+\log g_{2}^{2}\right)\right]\right)
\]
\[
\mathcal{H}_{g}[F_{k}(\mathbf{g})]=\left(\begin{array}{cc}
G_{20} & G_{11}\\
G_{11} & G_{02}
\end{array}\right)=\left(\begin{array}{cc}
g_{2}^{2}\left(H_{C}^{''}-f^{''}\log g_{2}^{2}\right) & 2g_{2}\left[H_{C}^{'}-f^{'}\left(1+\log g_{2}^{2}\right)\right]\\
2g_{2}\left[H_{C}^{'}-f^{'}\left(1+\log g_{2}^{2}\right)\right] & 2\left[H_{C}-f\left(3+\log g_{2}^{2}\right)\right]
\end{array}\right)
\]
\end{widetext}
where the primes denote derivation of functions of one variable with
respect to their argument. For brevity, we denote $G_{nm}=\partial_{g1}^{n}\partial_{g2}^{m}F_{k}(g_{1},g_{2})$. 

The gradient in terms of the original coordinates, takes the form
\begin{eqnarray}
\nabla F_{k}&=&\frac{\partial F_{k}}{\partial x_{j}}  =  \sum_{l}\frac{\partial F_{k}}{\partial g_{l}}\frac{\partial g_{l}}{\partial x_{j}}\\
&=&\sum_{l}\nabla_{g}[F_{k}(\mathbf{g})]_{l}\mathcal{J}[\mathbf{g}(\mathbf{x})]_{lj}=\nabla_{g}[F(\mathbf{g})]\cdot\mathcal{J}[\mathbf{g}(\mathbf{x})]\nonumber \\
 & = & G_{10}\nabla g_{1}+G_{01}\nabla g_{2}\label{eq:gradF}
\end{eqnarray}
The Jacobian of the vector function $\mathbf{g}(\mathbf{x})$ is defined
as $\mathcal{J}[\mathbf{g}(\mathbf{x})]_{ij}=\frac{\partial g_{i}}{\partial x_{j}}$.
The Hessian is 
\begin{eqnarray*}
\mathcal{H} F_{k} & = & \mathcal{J}[\nabla F_{k}]=\left[\frac{\partial^{2}F_{k}}{\partial x_{i}\partial x_{j}}\right]_{ij}=\frac{\partial}{\partial x_{i}}\left(\sum_{l}\frac{\partial F_{k}}{\partial g_{l}}\frac{\partial g_{l}}{\partial x_{j}}\right)\\
 & = & \sum_{l}\left\{ \sum_{l'}\left(\frac{\partial F_{k}}{\partial g_{l}\partial g_{l'}}\frac{\partial g_{l'}}{\partial x_{i}}\frac{\partial g_{l}}{\partial x_{j}}\right)+\frac{\partial F_{k}}{\partial g_{l}}\frac{\partial^{2}g_{l}}{\partial x_{i}\partial x_{j}}\right\} \\
 & = & \sum_{l}\sum_{l'}\left(\mathcal{J}[\mathbf{g}(\mathbf{x})]_{lj}\mathcal{H}_{g}[F_{k}(\mathbf{g})]_{ll'}\mathcal{J}[\mathbf{g}(\mathbf{x})]_{l'j}\right)\\
 & &+\sum_{l}\nabla_{g}[F_{k}(\mathbf{g})]_{l}\mathcal{H}[g_{l}(\mathbf{x})]_{ij}\\
 & = & \mathcal{J}[\mathbf{g}(\mathbf{x})]^{T}\mathcal{H}_{g}[F_{k}(\mathbf{g})]\mathcal{J}[\mathbf{g}(\mathbf{x})]+\nabla_{g}[F_{k}(\mathbf{g})]\mathcal{H}[\mathbf{g}(\mathbf{x})]\\
 & = & G_{20}\nabla g_{1}\otimes\nabla g_{1}+G_{02}\nabla g_{2}\otimes\nabla g_{2}\\
 & & +G_{11}\left(\nabla g_{2}\otimes\nabla g_{1}+\nabla g_{1}\otimes\nabla g_{2}\right)\\
 &  & +G_{10}\mathcal{H}(g_{1})+G_{01}\mathcal{H}(g_{2})
\end{eqnarray*}
Explicitly, 
\begin{eqnarray*}
[\nabla g_{1}]_{i} & = & \frac{\partial}{\partial x_{i}}r=\frac{\partial}{\partial x_{i}}\sqrt{\sum_{j}x_{j}^{2}}=\frac{x_{i}}{r}\\
{}[\nabla g_{2}]_{i} & = & \frac{\partial}{\partial x_{i}}\frac{x_{k}}{r}=\frac{1}{r^{3}}\left(r^{2}\delta_{ik}-x_{k}x_{i}\right)
\end{eqnarray*}
These two vectors are othogonal: another good feature of decomposing
$F$ in the form of Eq. (\ref{eq:F alt}). 

From Eq.(\ref{eq:gradF}) it is immediate to verify that $\partial_{x_{j}}[F_{k}(\mathbf{x})]\geq0$,
$\forall j$, because $G_{10}$ and $G_{01}$ are both non negative
in $\mathcal{D}$ as well as each component of $\nabla g_{1}$ and
$\nabla g_{2}$. Since each component of the gradient $\nabla[F(\mathbf{x})]$
is a sum of positive contributions, it follows that $F(\mathbf{x})$
is an increasing function with respect to each component of $\mathbf{x}$. 

The outer products of derivative terms are 
\begin{eqnarray*}
[\nabla g_{1}\otimes\nabla g_{1}]_{ij} & = & \frac{x_{i}x_{j}}{r^{2}}\\
{}[\nabla g_{2}\otimes\nabla g_{1}]_{ij} & = & \frac{1}{r^{4}}\left(r^{2}\delta_{ik}-x_{k}x_{i}\right)x_{j}=[\nabla g_{1}\otimes\nabla g_{2}]_{ji}\\
{}[\nabla g_{2}\otimes\nabla g_{2}]_{ij} & = & \frac{1}{r^{6}}\left(r^{2}\delta_{ik}-x_{k}x_{i}\right)\left(r^{2}\delta_{jk}-x_{k}x_{j}\right)
\end{eqnarray*}
and 
\begin{eqnarray*}
[\mathcal{H}g_{1}]_{ij} & = & \mathcal{J}[\nabla g_{1}]_{ij}=\frac{\delta_{ij}}{r}-\frac{x_{i}x_{j}}{r^{3}}=\frac{1}{r}\left(\mathbb{I}-\nabla g_{1}
\otimes\nabla g_{1}\right)\\
{}[\mathcal{H}g_{2}]_{ij} & = & \mathcal{J}[\nabla g_{2}]_{ij}
=\frac{\partial}{\partial x_{j}}\left[\frac{1}{r^{3}}\left(r^{2}\delta_{ik}-x_{k}x_{i}\right)\right]\\
 & = & -\frac{1}{r}\left(\nabla g_{2}\otimes\nabla g_{1}+\nabla g_{1}\otimes\nabla g_{2}\right)\\
 & & -\frac{x_{k}}{r^{3}}\left(\mathbb{I}-\nabla g_{1}\otimes\nabla g_{1}\right)
\end{eqnarray*}
Thus the Hessian becomes
\begin{eqnarray*}
\mathcal{H}F_{k} & = & \frac{1}{r^{3}}\left(x_{k}G_{01}-r^{2}G_{10}+r^{3}G_{20}\right)\nabla g_{1}\otimes\nabla g_{1}\\
 &  & +\left(G_{11}-\frac{1}{r}G_{01}\right)\left(\nabla g_{2}\otimes\nabla g_{1}+\nabla g_{1}\otimes\nabla g_{2}\right)\\
 && +G_{02}\nabla g_{2}\otimes\nabla g_{2}+\frac{1}{r^{3}}\left(r^{2}G_{10}-x_{k}G_{01}\right)\mathbb{I}
\end{eqnarray*}
This matrix can be decomposed in direct sum of a non singular $2\times2$
matrix written in the basis $\{\nabla g_{1},\nabla g_{2}\}$ and a
uniform diagonal part. The square norm of $\nabla g_{2}$ is 
\[
\Vert\nabla g_{2}\Vert^{2}=\frac{1}{r^{4}}\left(r^{2}-x_{k}^{2}\right)=\frac{1}{r^{2}}\left(1-g_{2}^{2}\right)
\]
Introducing the orthonormal basis $b_{1}=\nabla g_{1}$ and $b_{2}=\nabla g_{2}/\Vert\nabla g_{2}\Vert$,
we obtain
\begin{eqnarray*}
\mathcal{H}F_{k} & = & \frac{1}{r^{2}}\left(g_{2}G_{01}-rG_{10}+r^{2}G_{20}\right)b_{1}\otimes b_{1}\\
 & & +\frac{1-g_{2}^{2}}{r}\left(G_{11}-\frac{1}{r}G_{01}\right)\left(b_{1}\otimes b_{2}+b_{2}\otimes b_{1}\right)\\
 & & +\frac{1-g_{2}^{2}}{r^{2}}G_{02}b_{2}\otimes b_{2}+\frac{1}{r^{2}}\left(rG_{10}-g_{2}G_{01}\right)\mathbb{I}
\end{eqnarray*}

In the two dimensional subspace spanned by $b_{1}$ and $b_{2}$,
the Hessian becomes 
\[
\mathcal{H}_{\parallel}F_{k}=\left(\begin{array}{cc}
\alpha & \beta\\
\beta & \gamma
\end{array}\right)
\]
where 
\begin{eqnarray*}
\alpha & = & G_{20}\\
\beta & = & \frac{\sqrt{1-g_{2}^{2}}}{r}\left(G_{11}-\frac{1}{r}G_{01}\right)\\
\gamma & = & \frac{1-g_{2}^{2}}{r^{2}}G_{02}+\frac{1}{r^{2}}\left(rG_{10}-g_{2}G_{01}\right)
\end{eqnarray*}
In the complementary space, we simply have 
\[
\mathcal{H}_{\perp}F_{k}=\eta(\mathbb{I}-b_{1}\otimes b_{1}-b_{2}\otimes b_{2})
\]
with $\eta=\frac{1}{r^{2}}\left(rG_{10}-g_{2}G_{01}\right)$. 

Using the Sylvester's criterion for testing the positivity of matrices,
we conclude that $F_{k}$ is convex if $\eta$, $\alpha$, and $\alpha\gamma-\beta^{2}$
are all non negative. These are continuous functions of two variables
defined in the set $r\in(0,1/2)$ and $g_{2}\in(0,1]$, so their sign
can be evaluated throuh standard calculus methods. The matrix element
$\alpha=g_{2}^{2}\left(H_{C}^{''}-f^{''}\log g_{2}^{2}\right)$ is
easily verified to be non-negative since $-g_{2}^{2}\log g_{2}^{2}\geq0$
and both $H_{C}$ and $f$ are convex functions. The quantity
\[
\eta=\frac{g_{2}^{2}}{r^{2}}\left(rH_{C}^{'}-2H_{C}+2f\right)-\frac{g_{2}^{2}}{r^{2}}\log g_{2}^{2}\left(rf^{'}-2f\right)
\]
is made up of two contributions, both positive; in fact $ $$rf^{'}-2f\geq0$
and $rH_{C}^{'}-2H_{C}+2f\geq0$. This latter inequality requires
some analysis. As a matter of fact, we can define the function $p(z)=(2z-1)(rH_{C}^{'}-2H_{C}+2f)$
and we study it as a function of $z\in[1/2,1]$, where $r^{2}=z(1-z)$.
We find that $p^{''}(z)=0$ has a single root in $z_{1}$ and is increasing:
this means that $p^{'}(z)$ has a negative minimum in $z_{1}$ and
has a zero in $z_{2}<z_{1}$ and $z_{3}=1$. Going back to $p(z)$,
we learn that it has a positive maximum in $z_{2}$ and $p(1/2)=p(1)=0$.
This means that $p(z)$ is nonnegative in its domain, hence also the
function $rH_{C}^{'}-2H_{C}+2f$. This proves that $\eta\geq0$. 

\begin{figure}
\begin{centering}
\includegraphics[scale=0.68]{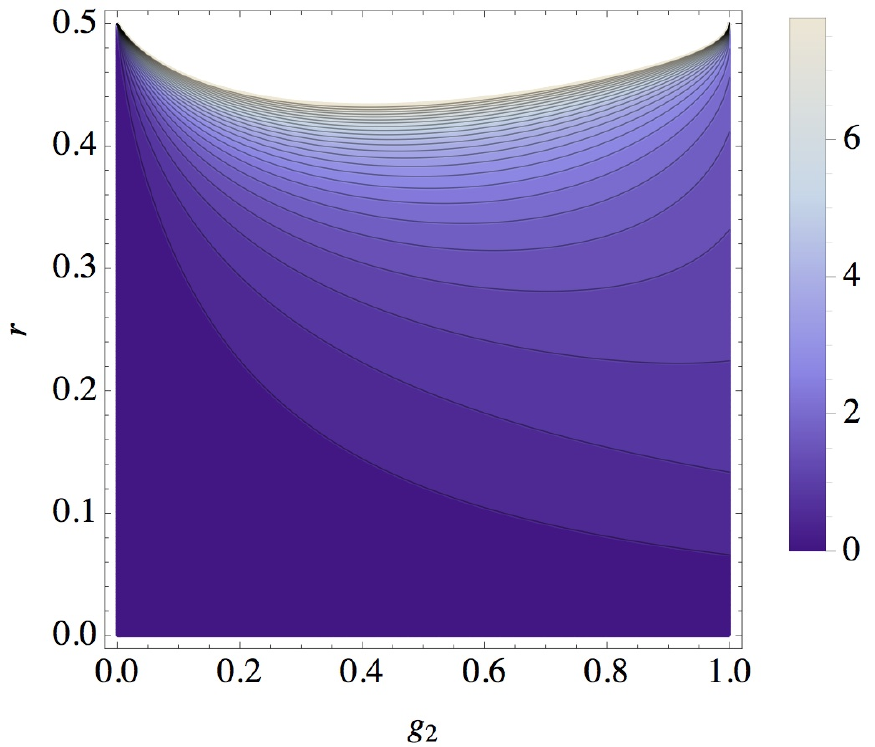}
\par\end{centering}

\caption{(Color online) The determinant $\alpha\gamma-\beta^{2}$ can be proven to be non
negative as a function of its explicit dependence $r^{2}$ and $g_{2}^{2}$.
\label{fig:det}}
\end{figure}

Finally, the determinant
\begin{eqnarray*}
\alpha\gamma-\beta^{2} & = & \frac{1}{r^{2}}\left\{ \left(1-g_{2}^{2}\right)G_{20}G_{02}+G_{20}\left(rG_{10}-g_{2}G_{01}\right)\phantom{\left(\frac{0}{0}\right)^{2}}\right.\\
& &\left. -\left(1-g_{2}^{2}\right)\left(G_{11}-\frac{1}{r}G_{01}\right)^{2}\right\} 
\end{eqnarray*}
can be efficiently evaluated numerically with arbitrary precision,
as it is a continuous function of two variables in the compact domain
$r\in(0,1/2)$ and $g_{2}\in(0,1]$. An analytic proof in closed form
can be performed, in the same fashion as the previous quantities,
but it turns out to be rather cumbersome since it involves higher
derivatives and very long expressions. In Fig. \ref{fig:det} we show
graphically that $\alpha\gamma-\beta^{2}\geq0$ and in particular
we observe that it is monotonically increasing with $r$. $\square$ 

As this proof is valid for every permitted value of $r$ and $x_{k}$,
it is shown that every $F_{k}$ in the sum of Eq.(\ref{eq:F alt})
is convex. This concludes our proof of convexity of $F(\mathbf{x})$
in Eq.(\ref{eq:F(x)}). 

\section{Test on two-mode squeezed states}\label{TWB}

Given a Fock space of two bosonic modes $|n_{1}\rangle_{A}\otimes|n_{2}\rangle_{B}$,
the pure two-mode squeezed states have the form 
\begin{equation}
|\Psi_r\rangle=\frac{1}{\cosh(r)}\sum_{n=0}^{\infty}\tanh^{n}(r)\:|n\rangle_{A}\otimes|n\rangle_{B}\label{eq:TWB}
\end{equation}
where $ $$r\geq0$ is the squeezing parameter. Pair states of the
form (\ref{eq:TWB}) belong to an infinite dimensional Hilbert space
and are a good playground for testing lower bounds to the entanglement
of formation (EOF). Furthermore, they are exceptionally interesting
for experimental realizations in quantum technologies. 

The density matrix $\rho_{ij}=\langle j|\Psi_r\rangle\langle\Psi_r|i\rangle$ 
in the pair basis $|i\rangle\equiv|i,i\rangle$ is written as
\[
\rho_{ij}=|c_{i}c_{j}|=\frac{1}{\cosh^{2}(r)}\tanh^{i}(r)\tanh^{j}(r).
\]
The negativity for this state can be easily computed, giving $\mathcal{N}(\Psi_r) = e^{r}\sinh(r)$, and is finite for every finite $r$.  
On the one hand, the entropy can be computed exactly, yielding  
\begin{eqnarray*}
S(\Psi_r) & = & -\sum_{n=0}^{\infty}c_{n}^{2}\log c_{n}^{2}\\
& = & \cosh^{2}(r)\log\left[\cosh^{2}(r)\right]-\sinh^{2}(r)\log\left[\sinh^{2}(r)\right]
\end{eqnarray*}

\begin{figure}
\centering{}\includegraphics[scale=0.4]{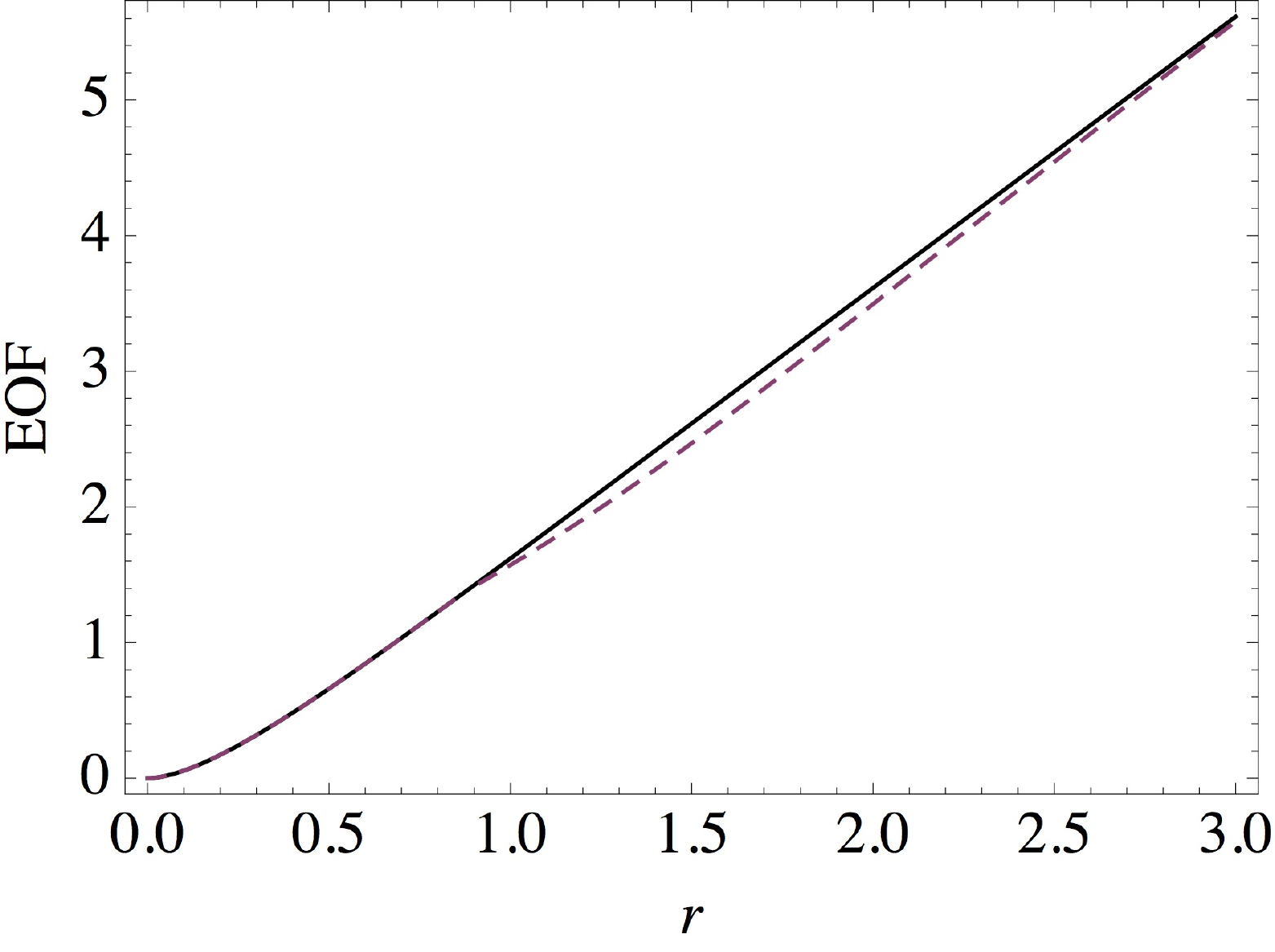}
\caption{(Color online) The exact entropy $S(\Psi_r)$ for the two-mode squeezed state (\ref{eq:TWB})
compared with the bound $F(\Psi_r)$ for the entanglement of formation.
\label{fig:EOF} }
\end{figure}

On the other hand our lower bound $F$ is given by 
\begin{eqnarray*}
F(\Psi_r) & = & \cosh^{2}(r)\log\left[\cosh^{2}(r)\right]-\sinh^{2}(r)\log\left[\sinh^{2}(r)\right]\\
 & + &  \Theta\left(\frac{1}{2}-\frac{1}{\cosh^{2}(r)}\right)\left\{ \cosh^{-2}(r)\log\left[\cosh^{-2}(r)\right]\right. \\
 & - &  \left. \tanh^{2}(r)\log\left[\tanh^{2}(r)\right]\right\} 
\end{eqnarray*}
where $\Theta(x)$ is the theta function of Heaviside. The lower bound
given in Ref.\cite{chen05} is $s(\mathcal{N})=0$, because in infinite
dimension we can always find states with finite negativity and zero
entropy. In Fig. (\ref{fig:EOF}) we can appreciate that in this test
case the lower bound $F$ to the EOF gives a very good estimation
of the entropy of formation (in this pure case, the entropy). 

\bibliographystyle{apsrev}

\end{document}